\documentclass[final,12pt,square]{elsarticle}

\usepackage{amssymb}
\usepackage{lineno}
\usepackage{soul}
\usepackage[colorlinks=true]{hyperref} 
\usepackage{color}
\usepackage{ulem}

\journal{NUCL INSTRUM METH A}

\begin{document}

\begin{frontmatter}


\title{Tritiated methane reduction in the PandaX-4T experiment via purge and cryogenic distillation processes}
\author[a,b]{Shuaijie Li}
\author[a,c,d,e]{Zhou Wang\corref{mycorrespondingauthor}} \ead{wangzhou0303@sjtu.edu.cn}
\author[c,d]{Xiangyi Cui}
\author[a,c,d,e]{Li Zhao\corref{mycorrespondingauthor}} \ead{zhaoli78@sjtu.edu.cn}
\author[f]{Yonglin Ju}
\author[a]{Wenbo Ma}
\author[g]{Yingjie Fan}
\author[a,c,d,e]{Jianglai Liu}
\author[b]{Liqiang Liu}
\author[b]{Kai Kang}
\cortext[mycorrespondingauthor]{Corresponding author}
\affiliation[a]{organization={School of Physics and Astronomy},
            addressline={Shanghai Jiao Tong University, Key Laboratory for Particle Astrophysics and
Cosmology (MoE), Shanghai Key Laboratory for Particle Physics and Cosmology},
            city={Shanghai},
            postcode={200240},
            country={China}}
            
\affiliation[b]{organization={Yalong River Hydropower Development Company, Ltd.},
            city={Chengdu},
            postcode={610051},
            country={China}}

\affiliation[c]{organization={New Cornerstone Science Laboratory, Tsung-Dao Lee Institute},
            addressline={Shanghai Jiao Tong University},             city={Shanghai},
             postcode={200240},
             country={China}}

\affiliation[d]{organization={Shanghai Jiao Tong University Sichuan Research Institute},
            city={Chengdu},
             postcode={610213},
             country={China}}

\affiliation[e]{organization={Jinping Deep Underground Frontier Science and Dark Matter Key Laboratory of Sichuan Province},
             country={China}}

\affiliation[f]{organization={Institute of Refrigeration and Cryogenic},
            addressline={Shanghai Jiao Tong University},             city={Shanghai},
             postcode={200240},
             country={China}}

\affiliation[g]{organization={Department of Physics},
            addressline={Yantai University},             city={Yantai},
             postcode={264005},
             country={China}}

\begin{abstract}
Tritium from tritiated methane (CH$_3$T) calibration is a significant impurity that restricts the sensitivity of the PandaX-4T dark matter detection experiment in the low-energy region. The CH$_3$T removal is essential for PandaX-4T and other liquid xenon dark matter direct detection experiments, as CH$_3$T serves as a critical component for low-energy calibration. To eliminate CH$_3$T, the xenon in the detector is suitably recuperated, leaving 1.8 bar of xenon gas inside, and the detector is flushed with heated xenon gas. Concurrently, leveraging the lower boiling point of methane relative to xenon, the PandaX-4T cryogenic distillation system is effectively utilized to extract CH$_3$T from xenon after optimizing the operational parameters. Following the commissioning run, 5.7 tons of xenon are purified via the distillation method. Recent data indicate that the CH$_3$T concentration reduces from $3.6\times10^{-24}$ mol/mol to $5.9\times10^{-25}$ mol/mol, demonstrating that gas purging and distillation are effective in removing CH$_3$T, even at concentrations on the order of $10^{-24}$ mol/mol.
\end{abstract}


\end{frontmatter}


\section{Introduction}
\label{Introduction}
Based on extensive astrophysical observations from cosmology and astronomy, it is established that a non-luminous, massive component of the universe, termed dark matter, exists~\cite{Bertone:2004pz},\cite{Roszkowski:2017nbc}. 
Among the various candidates for dark matter, Weakly Interacting Massive Particles (WIMPs) are considered the most promising. 
PandaX-4T, XENONnT and LUX-ZEPLIN (LZ) are leading the direct detection experiments using liquid xenon to search for WIMPs~\cite{PandaX:2024qfu},\cite{XENON:2023cxc},\cite{LZ:2022lsv}. 
Liquid xenon is mainly favored as a detection medium due to its high scintillation light yield, low energy threshold, substantial nuclear mass, which enhances WIMP-nucleus interactions and low intrinsic radioactive background~\cite{Liu:2017drf}.
\par
The PandaX-4T experiment, situated at the China Jinping Underground Laboratory (CJPL), employs a dual-phase xenon time projection chamber (TPC) to detect dark matter particles~\cite{PandaX-4T:2021bab},\cite{Kang:2010zza},\cite{Li:2014rca}.
Building on the successes of the PandaX-I and PandaX-II experiments over the past decade~\cite{PandaX:2015gpz},\cite{PandaX-II:2020oim}, PandaX-4T represents the latest generation of this experimental series, utilizing a total of 5.7 tons of xenon, with a target mass of 3.7 tons of liquid xenon in the TPC. The results from its commissioning run were appropriately documented in Refs.~\cite{PandaX:2024qfu},\cite{PandaX-4T:2021bab}. 
Achieving an extremely low background is crucial for enhancing the sensitivity of dark matter detectors. The primary intrinsic radioactive contaminants in the xenon medium are $^{85}$Kr and $^{222}$Rn~\cite{PandaX-4T:2021bab},\cite{XENON:2018voc},\cite{LUX:2016ggv},\cite{Ahlswede:2021jer}.

\par
Cryogenic distillation, which takes advantage of the differences in boiling points among xenon (165 K), krypton (120 K), and radon (211 K) at atmospheric pressure, serves as an effective technique for purifying xenon by removing krypton and radon. For instance, the XMASS and XENON1T experiments have developed cryogenic distillation systems capable of reducing krypton concentration by five orders of magnitude in a single pass~\cite{Wang:2014ehv},~\cite{XENON:2016bmq}. 
In comparison, the cryogenic distillation system employed by PandaX-4T successfully lowered the krypton concentration from 0.5 ppm (part-per-million) to 0.3 ppt (part-per-trillion), achieving a reduction factor of $1.7\times10^{6}$~\cite{PandaX-4T:2021bab},\cite{Cui:2020bwf}.
\par
In addition to $^{85}$Kr and $^{222}$Rn, tritium emerged as another significant radioactive contaminant that negatively impacted the background during the commissioning run of PandaX-4T for dark matter detection~\cite{PandaX-4T:2021bab}. 
The tritium was a byproduct of the CH$_3$T calibration conducted at the end of the PandaX-II experiment. 
A total of 1.1 tons of xenon sourced experimentally from PandaX-II, containing CH$_3$T at $10^{-22}$ mol/mol, was distilled and subsequently introduced into the PandaX-4T detector before commissioning. 
The CH$_3$T calibration was initially employed in the LUX experiment to investigate the detector's electron-recoil (ER) response at lower energy levels~\cite{LUX:2015amk}. 
However, due to tritium's half-life of 12.3 years, it does not decay sufficiently during the operational timeline of the detector. One potential strategy for tritium removal involves substituting it with deuterium, as exemplified by the Darlington Tritium Removal Facility, which employs a replacement reaction between D$_2$ and DTO~\cite{Busigin:1987}. 
For addressing surface contamination, a method combining ultraviolet radiation and ozone gas has been developed~\cite{Dobi:2010ai}. An alternative approach utilizes specialized commercial getters operating at elevated temperatures: in the LUX experiment, a zirconium (Zr) getter heated to 450 $^\circ$C demonstrated effectiveness in reducing injected CH$_3$T concentrations~\cite{Dobi:2010ai}. 
Notably, when the removal efficiency of the getter reached a plateau during the PandaX-II CH$_3$T calibration, a method that integrated hot xenon gas flushing with cryogenic distillation was successful in lowering tritium concentration from $2.1\times10^{-22}$ mol/mol to $4.9\times10^{-24}$ mol/mol~\cite{PandaXII:2020udv}. 
Following a single offline distillation, the tritium concentration in xenon during the PandaX-4T commissioning run was measured at $6.6\times10^{-24}$ mol/mol. 
The experimental data from this commissioning run revealed the limitations of chemical adsorption methods at such ultra-low concentrations: the 400 $^\circ$C commercial Zr getters utilized in the PandaX-4T experiment attempted to mitigate the tritium contamination, but the reduction of CH$_3$T was relatively slow~\cite{PandaX-4T:2021bab}. 
Dobi et al.~\cite{Dobi:2010ai} examined the performance of the Zr getter for xenon gas purification, concluding that the Zr getter achieves optimal performance for methane at higher getter temperatures and lower flow rates-specifically, at 450 $^\circ$C and 2.5 standard liters per minute (SLPM), yielding a methane purification efficiency exceeding 99.99\%, which aligns with background levels. In contrast, at 400 $^\circ$C and 10 SLPM, the efficiency drops to 92.5\%, as illustrated in Figure 4 of Ref.~\cite{Dobi:2010ai}. 
As a result, the Zr getters employed by PandaX-4T at 400 $^\circ$C with a flow rate of 80 SLPM do not match the performance of those used in the LUX experiment, which operated at 450 $^\circ$C with the flow rate of 27 SLPM. 
\par
Following the commissioning run of PandaX-4T, the liquid xenon in the detector was successfully recuperated, leaving a residual pressure of 1.8 bar of xenon gas. 
The detector was then flushed with hot xenon gas, which was heated to approximately 80 $^\circ$C via electric heating belts. 
Employing a cryogenic distillation technique, the more volatile CH$_3$T component-characterized by a boiling point of 112 K-was effectively removed from the xenon. Over a period of 40 days of offline distillation, 5.7 tons of xenon from the inner vessel of the detector underwent distillation, achieving a tritium concentration reduction factor of 6. This outcome demonstrates that the combined approach of hot xenon gas flushing and cryogenic distillation is effective for tritium removal at ultralow concentrations ($10^{-24}$ mol/mol). 
In the PandaX-4T setup, tritium concentrations were reduced to 0.02 $\mu$Bq/kg ($3.6\times10^{-24}$ mol/mol) using Zr getters at 400 $^{\circ}$C, although under this operating condition, these getters were less effective at removing CH$_3$T. Continuous distillation further reduced the CH$_3$T concentration to 0.005 $\mu$Bq/kg ($5.9\times10^{-25}$ mol/mol). In the LUX experiment, thanks to the efficient Zr getters operating at 450 $^{\circ}$C and the use of a titanium cryostat, the tritium level reached approximately 0.002 $\mu$Bq/kg, with a target of <0.33 $\mu$Bq for 145 kg Xe~\cite{LUX:2015amk}. Neither XENON1T nor XENONnT employed CH$_3$T calibration, however, XENON100, which utilized similar Zr getters and detector materials as PandaX, faced a comparable challenge as PandaX-4T post-CH$_3$T calibration, with tritium levels reduced to around 150 $\mu$Bq/kg after four months of purification~\cite{Xenon100tritium}.
\par
This paper presents a novel cryogenic distillation methodology combined with xenon flushing techniques for the removal of tritium from xenon at ultralow concentrations. The xenon gas flushing process utilized within the detector is detailed in Sec. 2. Sec. 3 elaborates on the structural design and key operational parameters of the PandaX-4T cryogenic distillation system. The calculations and simulations relating to tritium removal efficiency, along with optimized operational parameters, are provided in Sec. 4. The analysis of experimental data and the resultant tritium concentrations are presented in Sec. 5, culminating in our conclusions and discussions in Sec. 6.

\section{Heated xenon gas flushing of the detector}
\label{section 2}
The PandaX-4T experiment inherited 1.1 tons of xenon from the PandaX-II detector. This xenon, which contained CH$_3$T, was used to flush the detector and the circulation system during the commissioning run. Residual CH$_3$T molecules that adhere to the detector surfaces and pipeline walls pose a limitation to the sensitivity of the detector. Common methods to resist physical adsorption encompass heating, decompressing, flushing and replacement~\cite{desorption}. 
In the PandaX-4T experiment, we utilized heating belts to elevate the temperature of the remaining xenon and employed zirconium-based getters within the circulation systems to effectively flush the detector. To assess the efficacy of flushing at elevated temperatures in mitigating potential outgassing from materials, the Arrhenius law is applied to describe the outgassing rate~\cite{arrl}:
\begin{equation}
k = A_0\cdot e^{-\frac{E_a}{k_BT}}\label{eq1} 
\end{equation}
where $A_0$ denotes the pre-exponential factor, $E_a$ represents the activation energy (eV), $k_B$ is the Boltzmann constant ($8.617\times10^{-5}$ eV/K), and T is the temperature. 
According to Eq.\ref{eq1}, considering that the ranges of $E_a$ for detector materials (stainless steel, polytetrafluoroethylene and copper) are from 0.3 to 0.7 keV, the calculated outgassing rate at 300 K is $1.8\times10^{3}$ to $3.7\times10^{7}$ times greater than the rate at 178 K. This result indicates that high-temperature flushing is an effective approach for mitigating outgassing.

\par
\begin{figure}[htbp]
 \centering\includegraphics[width=10cm, angle=0]{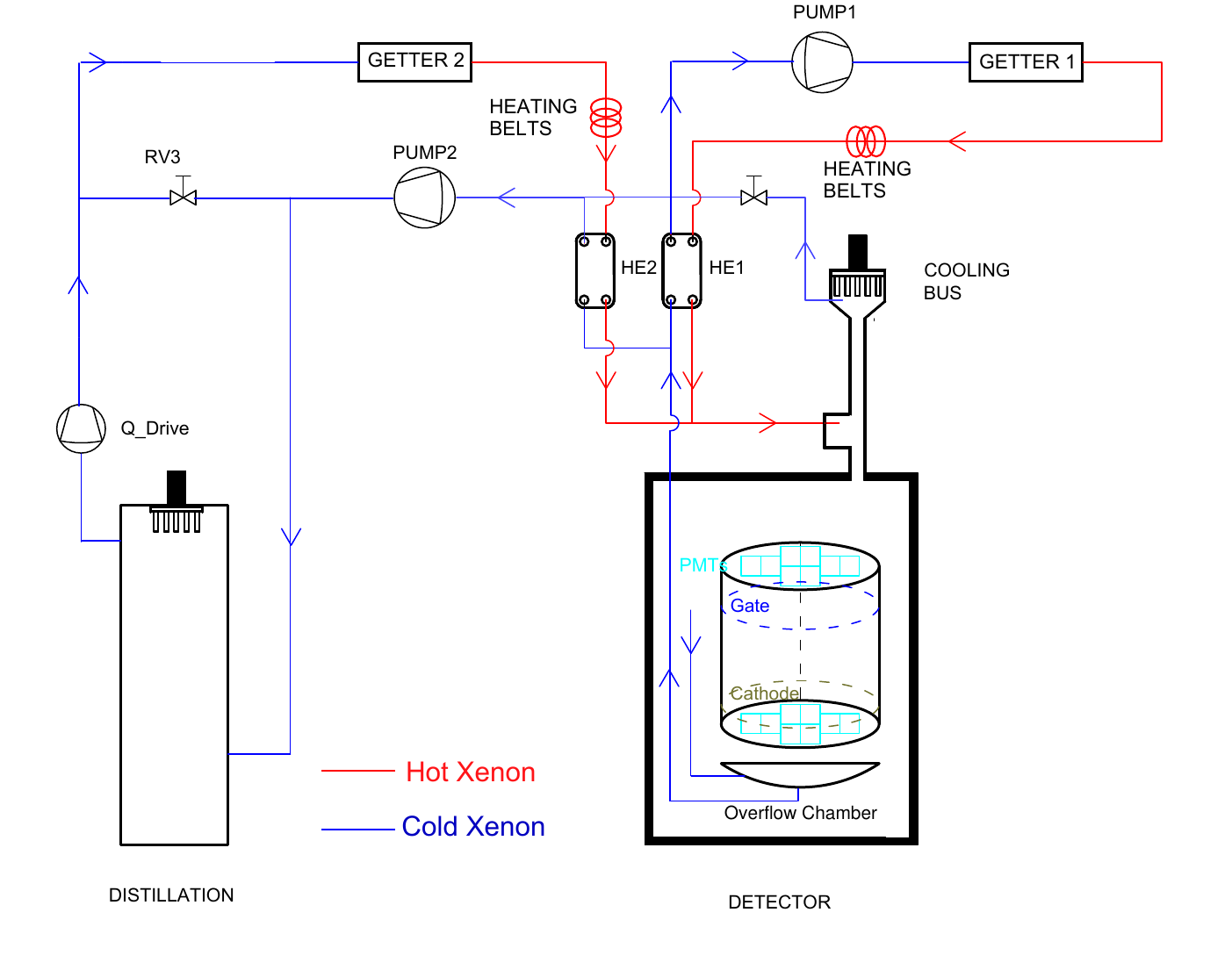}
 \caption{Flow diagram of the PandaX-4T cryogenic circulation system with heating belts.}
 \label{fig:layout}
 \end{figure}

As illustrated in the flow diagram in Fig.~\ref{fig:layout}, xenon from the detector was extracted using KNF pumps and subsequently purified via a commercial getter within the circulation system. The commercial zirconium-based getter (PS5-MGT50R-909 Rare Gas Purifier, SAES Inc.) operates at 400 $^{\circ}$C to effectively remove H$_2$O, O$_2$, CO, CO$_2$, H$_2$, N$_2$, and total hydrocarbons (THC) including CH$_4$, by forming irreversible chemical bonds.
\par
During the detector flushing, the pressure of the circulating xenon gas, heated by heating belts encircling the pipeline, was kept at about 1.8 bar.
The heating temperature was maintained at 80 $^{\circ}$C and the flow rate was approximately 80 SLPM while the getters were operational. This flushing process extended over a duration of more than one month, following which the detector underwent a vacuuming period of approximately one month. The vacuum of the inner vessel was kept at 0.05 Pa.

\section{PandaX-4T cryogenic distillation system setup and technology process}
\label{section 3}
The PandaX-4T cryogenic distillation system was originally developed for the removal of krypton and radon from xenon~\cite{Cui:2020bwf},\cite{yanrui:2021}. Given that CH$_3$T possesses a lower boiling point than krypton, this system can theoretically be adapted for the removal of tritium, as it shares a similar operational process with krypton distillation.

\subsection{System setup and equipment}
\label{sec:3:1}
    \begin{figure}[htbp]
    \centering\includegraphics[width=6cm, angle=0]{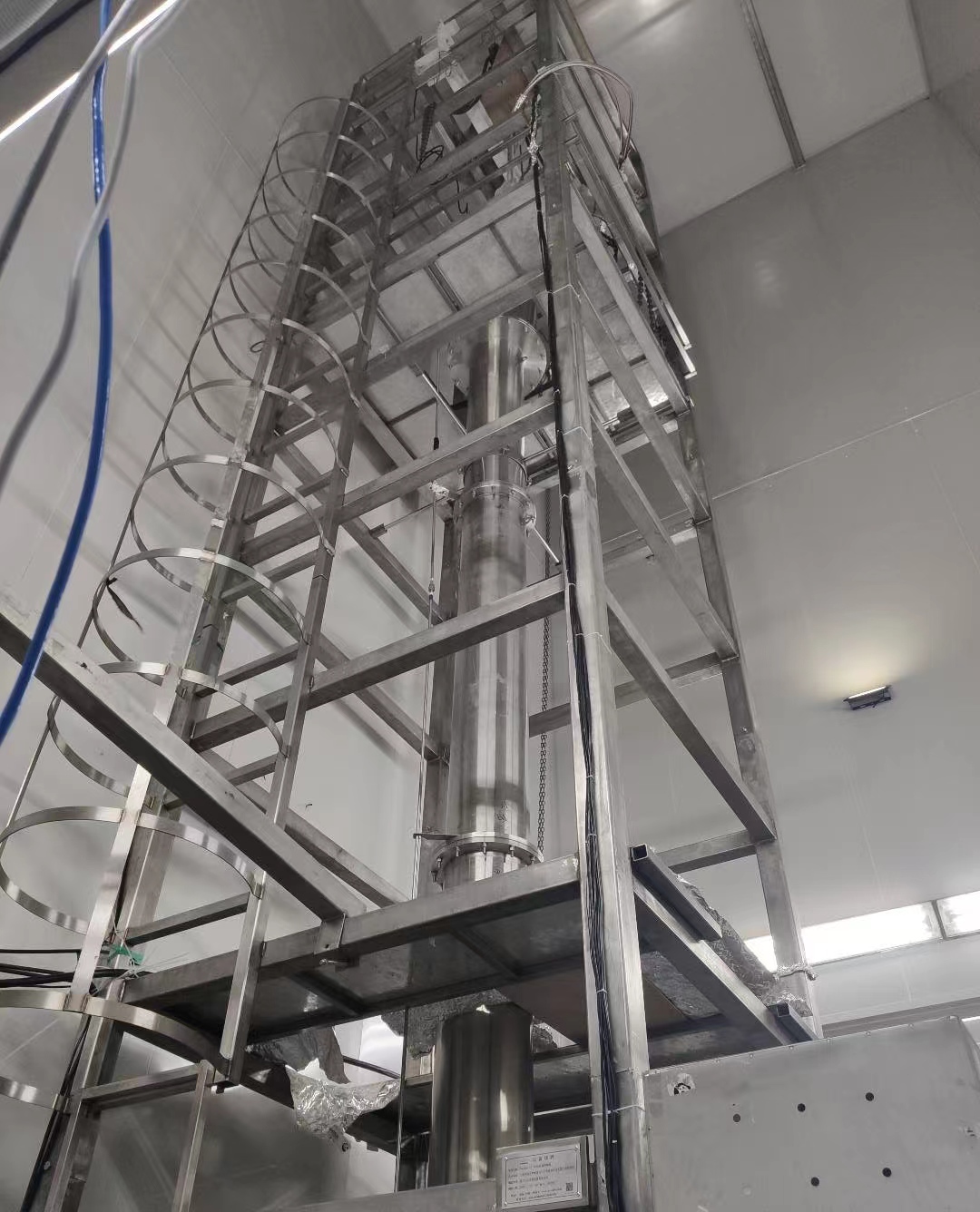}
    \caption{Photo of PandaX-4T distillation tower.}
    \label{fig.dt}
    \end{figure}
The photograph of the distillation system is presented in Fig.~\ref{fig.dt}. A key element of the distillation system is the packing column, which is filled with stainless-steel structured packing featuring a specific surface area of 1135 m$^{2}$/m$^{3}$. Heat and mass transfer between the gaseous and liquid xenon occurs at the packing, facilitating the attainment of gas-liquid equilibrium. At the top of the distillation tower is the condenser, which is coupled with a GM cryocooler (AL300, Cryomech Inc.) that delivers a maximum cooling power of 400 W to maintain a temperature of 181.5 K, thereby liquefying a portion of the ascending gaseous xenon and achieving a reflux ratio of 220. The liquid xenon descends into the reboiler through the packing column. The reboiler, situated at the bottom of the tower, is equipped with three heating rods, each providing a maximum power output of 180 W. During operation, approximately 30 kg of liquid xenon is stored in the reboiler, with a fraction vaporized by the heaters for recycling through the packing column. The cryogenic distillation tower is thermally insulated using multi-layer insulation and high vacuum technology, maintained at a pressure of $3\times10^{-3}$ Pa, resulting in a total heat leakage of less than 15 W~\cite{Cui:2020bwf},\cite{yanrui:2021}.


\subsection{Technology process}
\label{sec:3:2} 
\begin{figure}[htbp]
\centering\includegraphics[width=10cm, angle=90]{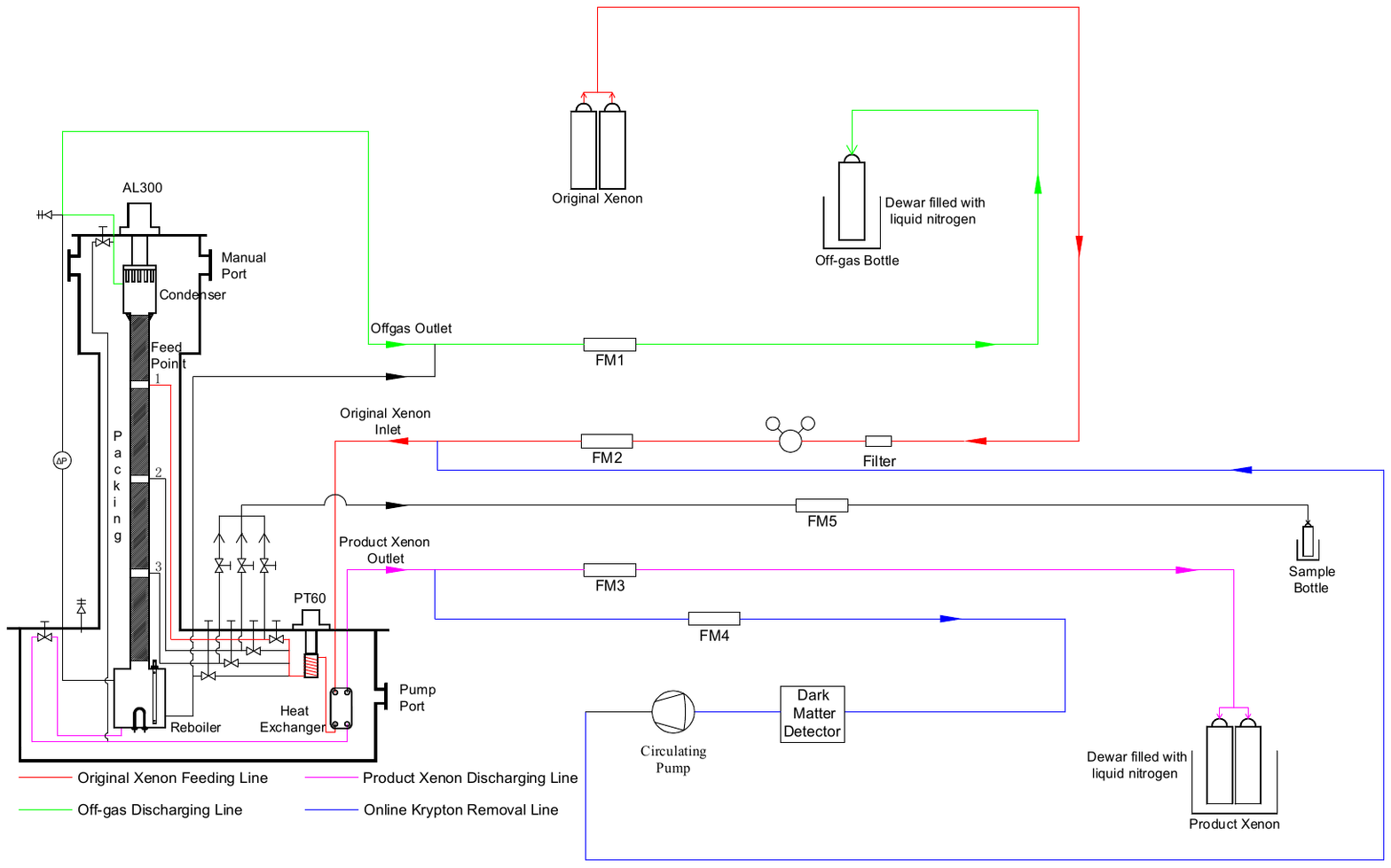}
\caption{Flow diagram of PandaX-4T distillation system.}
\label{fig.flowchart}
\end{figure}
The flow diagram of the distillation system is illustrated in Fig.~\ref{fig.flowchart}, which includes the feeding line, the off-gas discharge line, the product xenon discharge line, and the online distillation line, represented in red, green, magenta and blue, respectively. The feeding line connects the distillation system to the gas storage system, allowing the original xenon gas from the gas cylinders to flow through a regulator and a flow controller (FM2). This setup reduces the gas pressure from 60 bar to 2.3 bar while maintaining a xenon flow rate of 30 SLPM, corresponding to 10 kg/h. The xenon gas passes through a commercial getter with a hydrogen removal unit, effectively reducing the concentrations of O$_2$, N$_2$, CO$_2$, CO and H$_2$O, especially CH$_4$, H$_2$, deuterium, tritium, and other hydrides to below 1$\times10^{-9}$ mol/mol. Afterwards, the original xenon gas is pre-cooled by the liquid xenon product, which flows out from the reboiler, at a heat exchanger. Then the cold xenon gas is cooled to approximately 181.5 K by the pre-cooling cryocooler (Model PT60, Cryomech Inc.) and fed into the distillation column at the upper feeding point 1 shown in Fig.~\ref{fig.flowchart}. The reason for using this feeding point is described in Sec.~\ref{section 4}.

\par
After undergoing the heat and mass transfer processes between the liquid and gaseous xenon in the packings of the distillation column, the product xenon is extracted from the reboiler at the bottom of the inner distillation tower via pressure differential. It subsequently passes through a heat exchanger and a flow controller (FM4), which controls its flow rate at 29.7 SLPM (9.9 kg/h), before being directed to the dark matter detector. Furthermore, the off-gas is extracted from the condenser at the top of the inner distillation tower and transferred to an aluminum bottle, which is immersed in a dewar filled with liquid nitrogen. The off-gas flow rate is controlled by the flow controller (FM1) and is set at 0.3 SLPM (0.1 kg/h).

\section{Calculation and simulation of the CH$_3$T removal efficiency}
 \label{section 4}

The cryogenic distillation tower utilizes different boiling points to separate gas components. 
Since the distillation tower has been essentially designed for krypton and radon removal, the theoretical removal efficiency and operation parameters for CH$_3$T removal should be recalculated and optimized.
\par
\begin{figure}[htbp]
 \centering\includegraphics[width=8cm, angle=0]{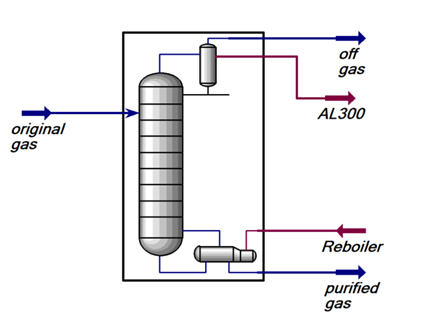}
 \caption{Schematically represented setup of the HYSYS simuation ( note: AL300 means the cooling power supplied by the cryocooler installed at the condenser).}
 \label{fig.sim1}
 \end{figure}

On the basis of the current distillation design, the critical operation parameters are simulated via Aspen HYSYS to confirm the optimized operation condition.
Aspen HYSYS is a professional chemical process simulation software that is able to simulate the geometrical and operational parameters of the distillation process. 
Due to the limitations of the software, the CH$_3$T concentration of the original xenon is set as $1.0\times10^{-10}$ mol/mol.
The simulation model is mainly constructed based on the design of the distillation tower, as illustrated in Fig.~\ref{fig.sim1}. 
The distillation column has a height of 6 m and an inner diameter of 125 mm. 
The number of theoretical plates is considered as 17. 
A theoretical plate in a distillation column is a hypothetical stage where the vapor and liquid streams reach perfect equilibrium. 
Heating power is specified at the reboiler, while the software computes the corresponding cooling power for the condenser. The reflux ratio and flow rates of the feed xenon gas, off-gas, and product xenon gas are considered user-defined inputs. 
For this purpose, the key operational parameters of the cryogenic distillation tower include the feeding points, reboiler heating power, reflux ratio, and feeding flow rate. 
The influences of pressure and temperature on CH$_3$T concentration in the purified xenon are limited to within one order of magnitude, allowing us to disregard pressure and temperature variations within the tower compared to other parameters~\cite{Cui:2020bwf},\cite{yanrui:2021}. 
The CH$_3$T concentration in the purified xenon is simulated under the following operational parameters:
\begin{enumerate}
  \item Based on the actual design of the PandaX-4T distillation tower, the feeding points of the distillation column are configured at the third, eighth and fourteenth theoretical plates counted from the top, designated as the middle-upper feeding point, middle feeding point and middle-lower feeding point, respectively.
  \item The reflux ratios of 145, 190 and 220 correspond to reboiler heating powers of 118 W, 154 W and 180 W, respectively.
  \item The feeding flow rates are set at 5 kg/h, 7.5 kg/h and 10 kg/h, respectively.
\end{enumerate}
  
\begin{figure}[htbp]
 \centering\includegraphics[width=15cm, angle=0]{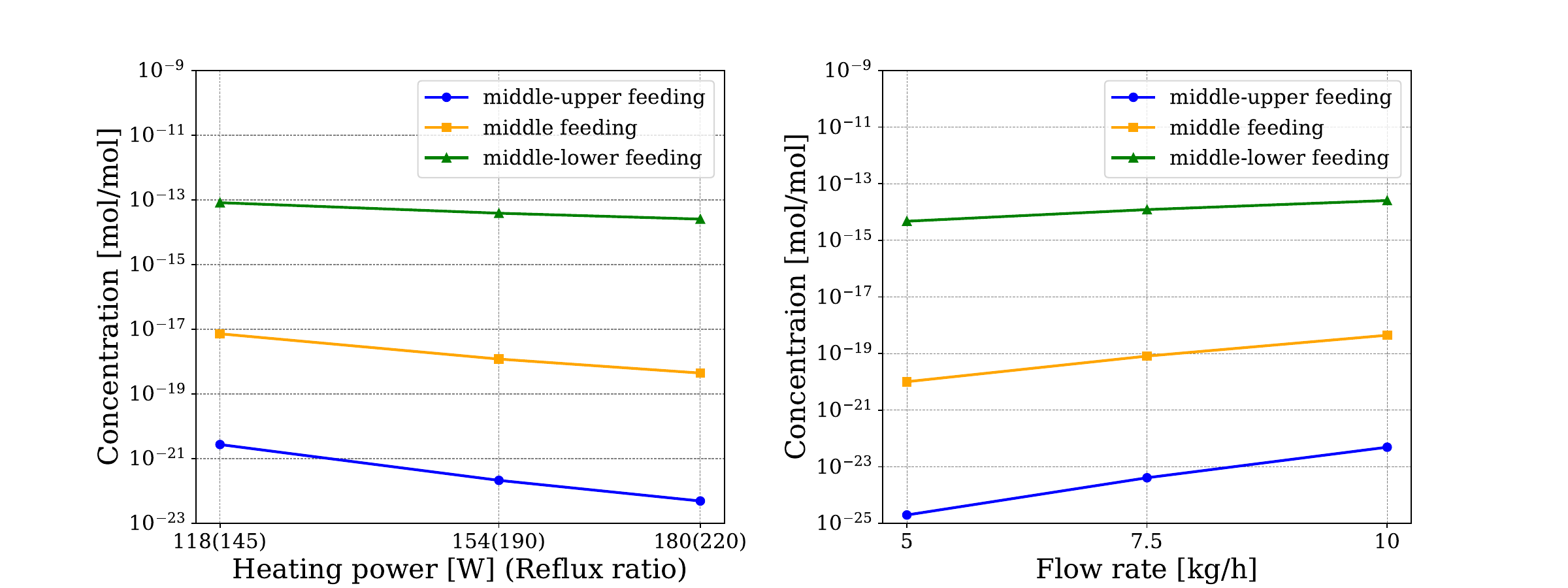}
 \caption{Left: Fixed flow rate at 10 kg/h with varying reboiler heating power and feeding points; Right: Fixed heating power at 180 W (reflux ratio = 220) with varying flow rate and feeding points.}
 \label{fig:simu_2}
 \end{figure}

The simulation results illustrating the impact of various operational parameters on the CH$_3$T concentration in the purified xenon are presented in Fig.~\ref{fig:simu_2}. 
These results indicate that the CH$_3$T concentration decreases as both the reflux ratio and reboiler heating power increase. 
Additionally, a lower flow rate leads to further reduction of CH$_3$T concentration.
When the initial CH$_3$T concentration in the original xenon is $1\times10^{-10}$ mol/mol, the purified xenon achieves a concentration as low as $2\times10^{-25}$ mol/mol, demonstrating exceptional purification efficiency. 
These results are derived from theoretical simulations based on the Soave-Redlich-Kwong (SRK) equation and the McCabe-Thiele (M-T) method. It is important to note that the M-T method is generally applied in typical chemical engineering contexts, where impurity concentrations are at the percentage level.
Currently, there is a lack of experimental data to confirm that the distillation tower retains effective removal efficiency at such low impurity concentrations. Additionally, in distillation processes dealing with extremely low concentrations, actual removal efficiency is significantly influenced by the specific operating conditions and the real gas-liquid equilibrium states. Nevertheless, the simulation results provide valuable insights into the optimal operating parameters of the distillation process.
 
Taking into account both CH$_3$T removal performance and distillation time, we selected the operational parameters of a middle-upper feeding point (3rd theoretical plate), a reboiler heating power of 180 W (reflux ratio of 220), and a flow rate of 10 kg/h. 
 
\section{Experimental results and data analysis}
\label{section 5}
The tritium radioactivity was identified as the dominant background signal during the commissioning run of the PandaX-4T experiment~\cite{PandaX-4T:2021bab}\cite{XENON:2020rca}, which concluded in April 2021. 
The xenon utilized in the detector was recuperated using a diaphragm pump and subsequently stored in 128 bottles (40 L each) under the pressure of 6 MPa~\cite{Wang:2022hkk}. 
A portion of the xenon gas, specifically 1.8 bar, was retained within the dark matter detector for flushing purposes, while the remaining xenon stored in the 128 bottles was subjected to a distillation process aimed at removing CH$_3$T.
 
 \subsection{Operation procedure}
  The distillation operation procedure comprised five distinct processes: pre-cooling, gas feeding, total reflux, purification and collection. 
 \begin{itemize}
 \item Pre-cooling Process \par
      Initially, the distillation tower was pressurized with xenon gas at 2.5 bar at ambient temperature. Cryocoolers (Model AL300 and PT60, as depicted in Fig.~\ref{fig.flowchart}) were activated to cool the inner tower to the target working temperature of 181.5 K. The AL300 cryocooler facilitated the liquefaction of xenon in the condenser, while the PT60 pre-cooled the feed xenon. To prevent pressure drops resulting from temperature reductions, xenon was intermittently introduced into the tower, maintaining the internal pressure between 1.8 and 2.5 bar. 
      As illustrated in Fig.~\ref{fig:ch4_pre}, the reboiler required approximately 6 hours to attain 181.5 K, indicating that the tower had been fully cooled and liquid xenon had formed in the reboiler, with the structured packings completely wetted. Upon completion of the pre-cooling phase, a continuous xenon feed into the inner tower commenced.
      
    \begin{figure}[htbp]
    \centering\includegraphics[width=10cm, angle=0]{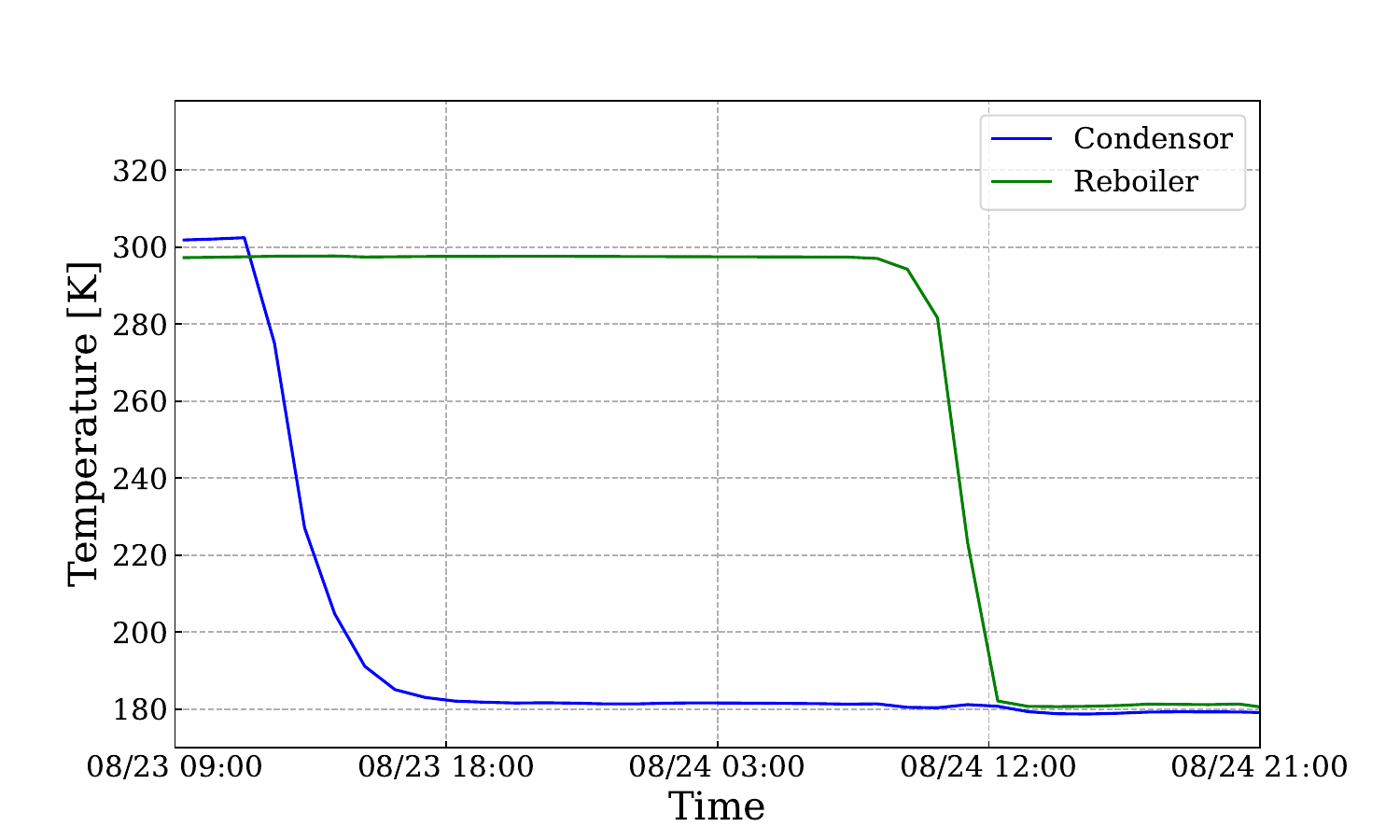}
    \caption{Temperature variations of the condenser (blue) and reboiler (green) during the pre-cooling process.}
    \label{fig:ch4_pre}
    \end{figure}
    
    \item Feeding process \par
     In this process, the original xenon gas was continuously fed into the distillation tower at a flow rate ranging from 5 to 20 SLPM until the liquid level in the reboiler reached approximately 20 cm after 6.5 hours, as shown in Fig.~\ref{fig:ch4_charge}. 
     Subsequently, the feeding of the original xenon gas was terminated.
     
    \begin{figure}[htbp]
    \centering\includegraphics[width=10cm, angle=0]{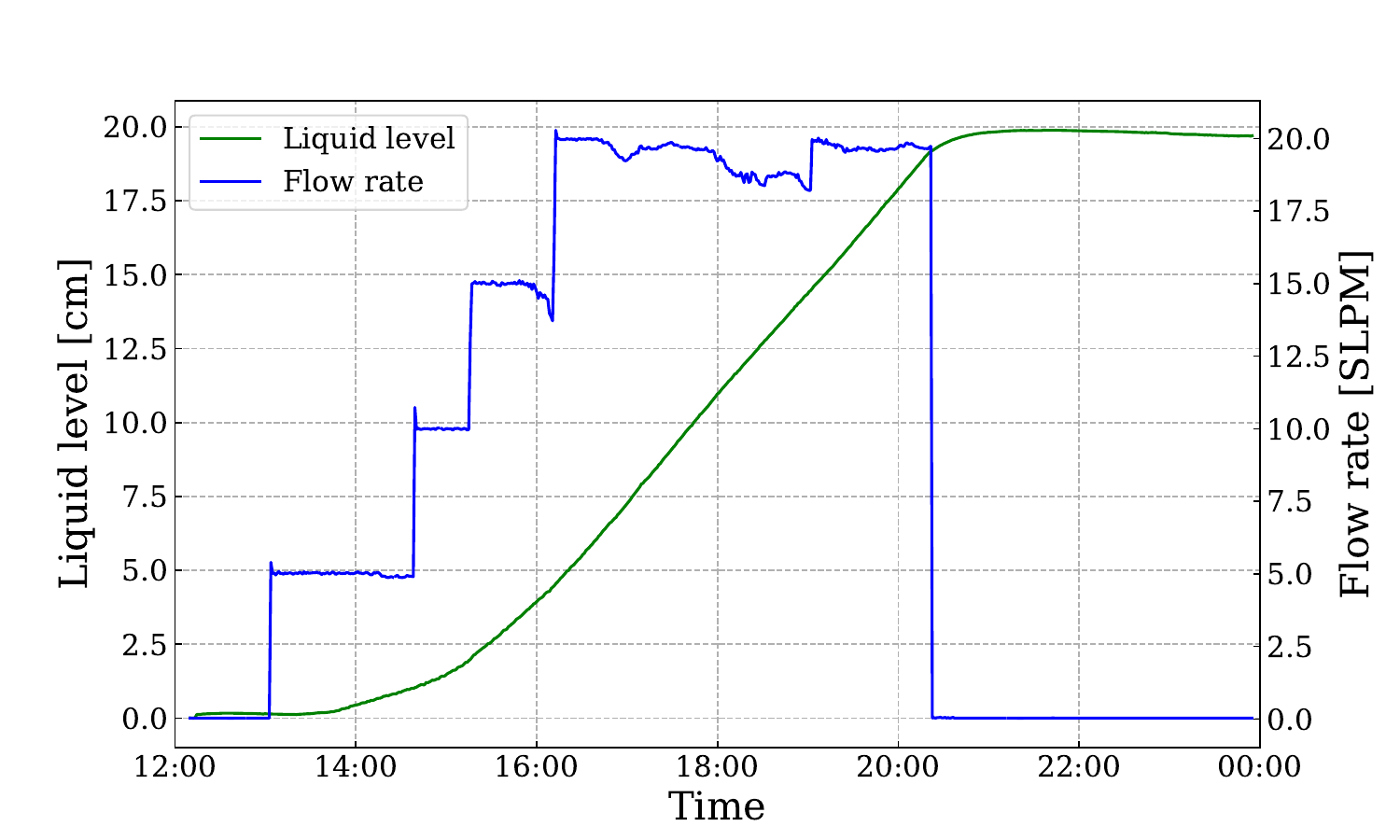}
    \caption{Liquid level (green) and flow rate (blue) variations during the gas feeding process.}
    \label{fig:ch4_charge}
    \end{figure}
 
    \item Total reflux process \par
    During this process, a heating power of 180 W was incrementally applied to the reboiler. 
    No xenon feed or discharge took place, allowing gas-liquid equilibrium to be established within the distillation tower through repeated evaporation-condensation cycles. 
    The total reflux process needed to be maintained for a minimum of 24 hours to ensure stabilized equilibrium. 
    Following this period, the reboiler contained liquid xenon with significantly reduced CH$_3$T concentrations, while the condenser accumulated the CH$_3$T-enriched off-gas.
    To evaluate distillation efficiency, xenon samples were extracted from the reboiler. The removal efficiency of CH$_3$T/Xe is inherently superior to that of Kr/Xe in principle due to the larger relative volatility, which is 15.3 versus 10.4, at the operating temperature of the distillation (178K)~\cite{Cui:2020bwf}. Given that CH$_3$T concentrations (at the level of $10^{-25}$ mol/mol) fall below the detection limit of the measurement system, krypton is utilized as an alternative indicator, with concentrations at approximately $10^{-12}$ mol/mol. During sample measurement, the eligible krypton concentration is needed to be at the ppt level within the sensitivity range of the krypton measurement system. 
    The absence of a detectable krypton signal in the product xenon during the measurement indicates that the distillation system was operating normally.
    \item Purification process \par
    The purification process is characterized as a dynamic distillation procedure. Original xenon was continuously supplied to the system at a flow rate of 30 SLPM. 
   Concurrently, product xenon and off-gas were discharged from the system at flow rates of 29.7 SLPM and 0.3 SLPM, respectively. 
    Throughout this process, product xenon samples were taken and analyzed. The results from the krypton measurement system indicated that the krypton concentration remained below the sensitivity threshold of the system. 
    Additionally, impurities with lower boiling points than xenon, such as O$_2$ and N$_2$ were remarkably reduced.
    This process spanned approximately one month, during which 5.7 tons of xenon were distilled. The experimental conditions are illustrated in Fig.~\ref{fig:ch_4_1}. 
    The data demonstrates that the reboiler temperature at the bottom was maintained at 181.28 $\pm$ 0.06 K, while the condenser temperature at the top stabilized at 183.13 $\pm$ 0.03 K. The condenser pressure exhibited slight fluctuations around 230 kPa. 
    The original xenon feed rate remained constant at 30 SLPM (10 kg/h), with purified xenon being discharged at 29.7 SLPM (9.9 kg/h) and off-gas vented at 0.3 SLPM (0.1 kg/h). 
    The liquid level in the reboiler was consistently maintained at approximately 20 cm. 
    These operational metrics confirm the thermodynamic stability of the distillation system throughout the extended purification process.
 \begin{figure}[htbp]
 \centering\includegraphics[width=10cm, angle=0]{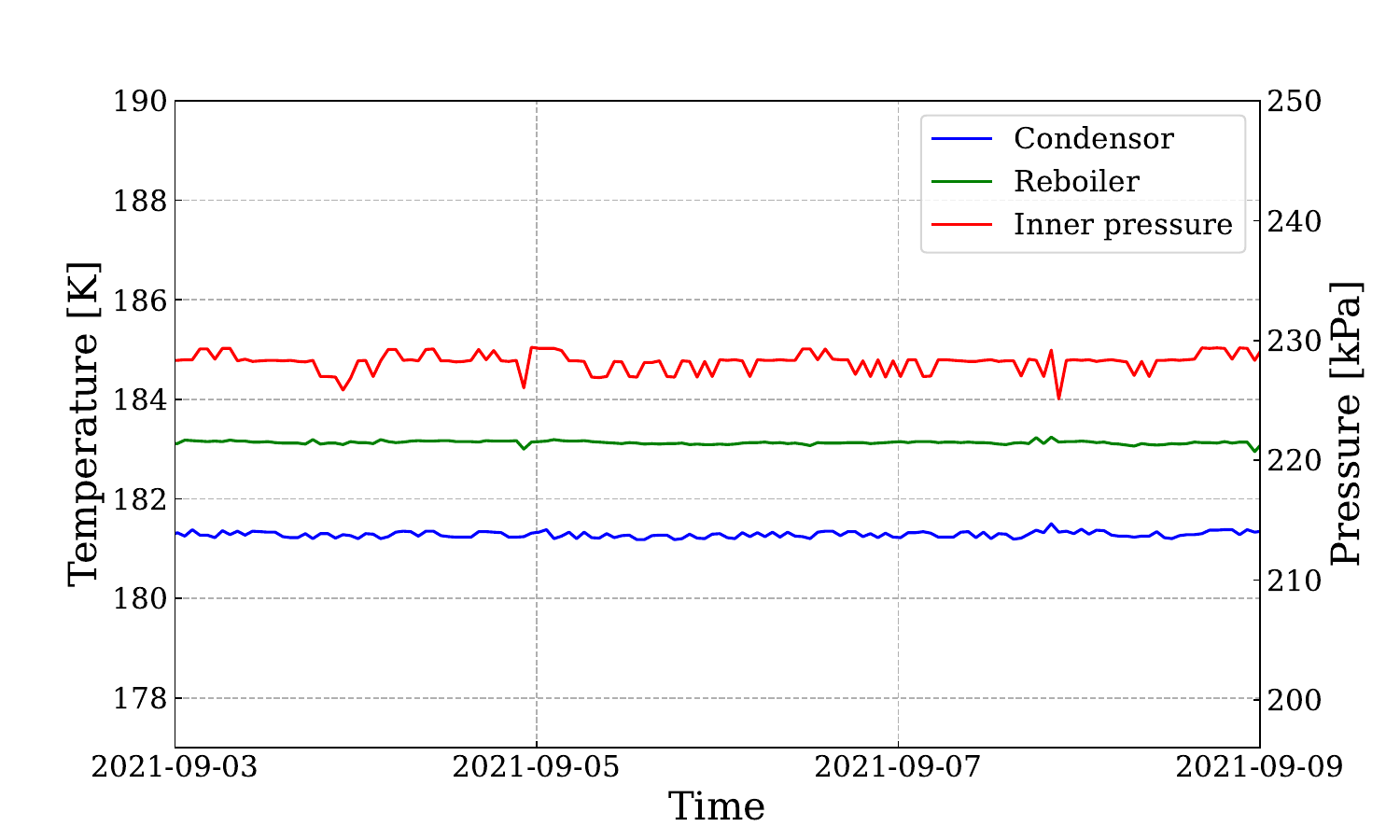}
 \caption{Temperature and pressure stability of the purification process.}
 \label{fig:ch_4_1}
 \end{figure}
     
     \item Collection process \par
     Once all xenon had been distilled and transferred into the detector, the system was deactivated, and residual xenon within the distillation column was collected.
     The product xenon and off-gas were then separately recuperated into two liquid nitrogen-cooled aluminum alloy bottles at a mass flow ratio of 99:1, which was maintained until the reboiler liquid level approached zero.
     Subsequently, the remaining xenon was collected as off-gas from the top of the distillation tower.

 \end{itemize}
    
\subsection{CH$_3$T concentration in PandaX-4T experiment}

Reviewing the history of the tritium issue in the PandaX experiments, the PandaX-II experiment first employed CH$_3$T calibration in July 2016, observing a peak CH$_3$T concentration of $2.1 \times 10^{-19}$ mol/mol. 
Following approximately four months of circulation through hot getters, the CH$_3$T concentration decreased to $2.1 \times 10^{-22}$ mol/mol and subsequently plateaued. 
A similar flushing and distillation process was subsequently conducted, resulting in a further decrease of the CH$_3$T concentration to $(4.9 \pm 1.2) \times 10^{-24}$ mol/mol. 
However, after data acquisition was completed, the PandaX-II experiment performed another CH$_3$T calibration, during which xenon was used for the PandaX-4T experiment. 
Before being filled into the PandaX-4T detector, the newly purchased xenon was mixed with the xenon from PandaX-II and distilled using the similar standard distillation method employed in PandaX-II, without optimizing the operational parameters for tritium removal. Consequently, tritium emerged as the dominant background in the commissioning run~\cite{PandaX-4T:2021bab}. 
\par
\begin{figure}[htbp]
 \centering\includegraphics[width=10cm, angle=0]{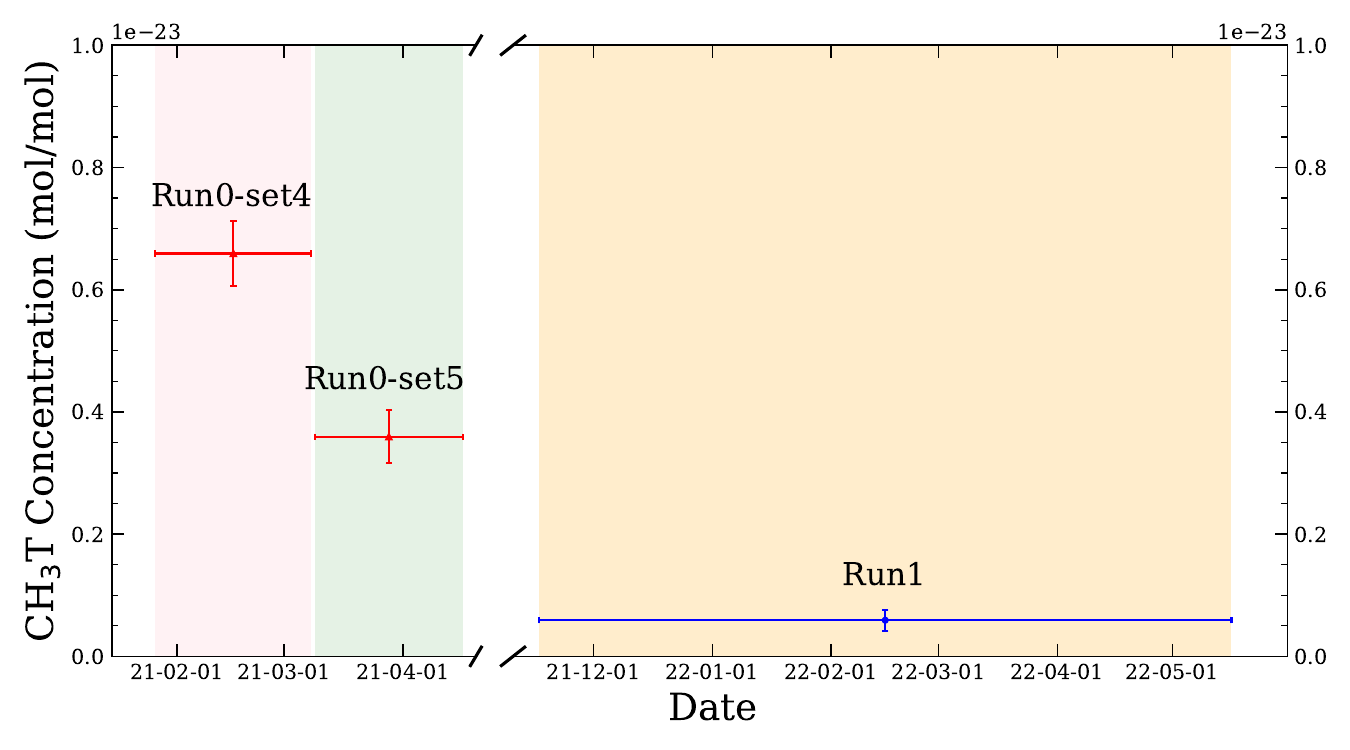}
 \caption{Tritium concentration evolution.}
 \label{fig:ch4t}
 \end{figure}

The CH$_3$T concentration at the $10^{-24}$ mol/mol level cannot be directly measured using conventional measurement systems. 
Nevertheless, the radioactivity of tritium generates ER signals in the TPC, which blend with other ER backgrounds, rendering them indistinguishable. 
The only method to quantify the CH$_3$T concentration involves fitting the energy spectrum. 
Through energy spectrum fitting of PandaX-4T experimental data, systematic variations in CH$_3$T concentration were observed between Run0 (commissioning run) and Run1, as illustrated in Fig.~\ref{fig:ch4t}. 
Notably, the getter reduced the CH$_3$T concentration from (6.6$~\pm~0.5) \times 10^{-24}$ mol/mol (Run0-set 4) to (3.6$~\pm~0.4) \times 10^{-24}$ mol/mol (Run0-set 5). 
Despite this reduction, tritium remained the dominant background component. 
The subsequent implementation of purge and cryogenic distillation processes with optimized operational parameters for tritium removal, conducted between Run0 and Run1, further decreased the CH$_3$T concentration to (5.9$~\pm~1.7) \times 10^{-25}$ mol/mol (Run1)~\cite{PandaX:2024qfu}.
\par
The ER event counts stand at 556 $\pm$ 33 for tritium and 504 $\pm$ 16 for others in Run0 (0.54 tonne $\cdot$ year), indicating that tritium dominates the ER background. Following the implementation of hot xenon gas flushing and distillation, the ER event counts are 114 $\pm$ 33 for tritium and 1226 $\pm$ 28 for other ER during Run1 (1 tonne $\cdot$ year), respectively~\cite{PandaX:2024qfu}. Consequently, the tritium background is no longer the predominant factor when compared to other ER backgrounds.
\par

\section{Conclusion and Discussion}
 \label{section 6}
Tritium represents one of the most significant backgrounds challenges in dark matter detection experiments. 
During the commissioning run of the PandaX-4T experiment, tritium was identified as the dominant background component.
Although the getter exhibited effectiveness in removing CH$_3$T, its practical implementation was constrained by prolonged processing times and suboptimal efficiency.
The combined approach of cryogenic distillation and hot xenon gas flushing resulted in substantial CH$_3$T removal. 
Analysis of tritium background components before and after the purge and distillation operations revealed an impressive purification efficiency of 84\%, reducing the concentration from $3.6\times10^{-24}$ mol/mol to 5.9$\times 10^{-25}$ mol/mol. As a result, tritium no longer dominates the ER background. These obtained results confirm that the integration of cryogenic distillation with xenon gas flushing is an effective method to reduce CH$_3$T contamination at the $10^{-24}$ mol/mol level, representing a low background advancement in ultra-trace impurity control for xenon-based detectors. 
 \par
Additionally, the experimentally obtained removal efficiency for CH$_3$T contamination is considerably lower than what was predicted by simulations based on M-T method. This discrepancy may result from several factors: (i) The M-T method is traditionally utilized in standard chemical engineering scenarios with impurity concentrations at the percentage level, which may not adequately reflect the purification capacity of the distillation column at extremely low concentrations; (ii) CH$_3$T molecules that adhere to the distillation tower packing could hinder the ideal gas-liquid exchange process; (iii) Trace amounts of CH$_3$T molecules that adhere to the surfaces of detector materials may not be entirely eliminated through purge processes. To further enhance tritium removal in the detector, online distillation with a high flow rate of 10 kg/h is currently under consideration.

\section*{Acknowledgments}
The authors would like to thank the supports of the PandaX collaboration. 

This project is supported by grants from National Key R\&D Program of China (2023YFA1606204, 2023YFA1606200), National Science Foundation of China (Nos.52206015, 12205189 and 12405220), grants from the Sichuan Science and Technology Program (No. 2024YFHZ0006), Sichuan Province Innovative Talent
Funding Project for Postdoctoral Fellows (No. BX202322), the Office of Science and Technology, Shanghai Municipal Government (Grant Nos. 23JC1410200, and 22JCJC1410200). We thank support from Double First Class Plan of the Shanghai Jiao Tong University, and the Tsung-Dao Lee Institute Experimental Platform Development Fund. We also thank the the sponsorship from the Chinese Academy of Sciences Center for Excellence in Particle Physics (CCEPP), Thomas and Linda Lau Family Foundation, Tencent and New Cornerstone Science Foundation in China, and Yangyang Development Fund. Finally, we thank the CJPL administration and the Yalong River Hydropower Development Company Ltd. for indispensable logistical support and other help.

\bibliographystyle{elsarticle-num} 
\bibliography{ref}

\end{document}